\documentclass[referee]{raa}            

\usepackage{graphicx,times}             
\usepackage{natbib}
\usepackage{amssymb,amsmath}
\bibpunct{(}{)}{;}{a}{}{,}

\usepackage[pagebackref=true]{hyperref}

\begin{document}

  \title{Calibrating Self-Weighted Shear Estimation via Field Distortion
}

   \volnopage{Vol.0 (20xx) No.0, 000--000}      
   \setcounter{page}{1}          

   \author{Zhenjie Liu 
      \inst{1,2}
   \and Jiarui Sun
      \inst{1}
   \and Jun Zhang
      \inst{1*}
   \and Cong Liu
      \inst{1}
   }

   \institute{State Key Laboratory of Dark Matter Physics, School of Physics and Astronomy, Shanghai Jiao Tong University, Shanghai 200240, China; {\it betajzhang@sjtu.edu.cn}\\
        \and
             Division of Physics and Astrophysical Science, Graduate School of Science, Nagoya University, Nagoya 464-8602, Japan\\
\vs\no
   {\small Received 20xx month day; accepted 20xx month day}}

\abstract{
Accurate cosmic shear measurement is the key to fully realize the scientific potential of large scale galaxy surveys. The Fourier\_Quad shear measurement method has been significantly developed to avoid biases caused by various factors, including the point spread function (PSF), photon noise, pixelation effect, selection effects, etc.. Shear statistics has also been optimized by using the PDF-SYM method (symmetrization of the Probability Distribution Function of the shear estimators) to achieve the minimal statistical error, without introducing systematic error. Nevertheless, the iterative nature of the PDF-SYM method makes it computationally expensive for massive datasets. To substantially improve efficiency, we introduce the Self-Weighted Shear Estimation (SWSE) method, which employs a specific self-weighting scheme to naturally suppress the shape noise as well as to balance the contributions from the bright and faint galaxies. Because SWSE is intrinsically biased, the primary focus of this work is to robustly evaluate whether the field-distortion test (FD test, a way to calibrate the shear bias on real data) can accurately recover and correct these inherent multiplicative and additive biases. Using both simulated and real survey data, we demonstrate that the FD test can successfully calibrate SWSE on galaxy-galaxy lensing measurements, enabling it to achieve statistical precision comparable to PDF-SYM. Our results establish the combination of FD test and SWSE as a robust shear estimation approach for forthcoming large-scale weak lensing surveys.
}

   \authorrunning{Liu et al. }            
   \titlerunning{SWSE \& FD test}  

   \maketitle

%
%

\section{Introduction}
Weak gravitational lensing serves as a crucial tool for probing the large-scale structure of the universe. It is caused by the slight bending of light paths by gravitational fields, causing subtle distortions (cosmic shear) in the shapes of distant galaxies. These distortions are only on the order of about one-tenth of the intrinsic shape of the galaxies, but they directly reflect the distribution features of dark matter halos, cosmic web,  and the distribution of dark matter \citep{Bartelmann:2017}. Therefore, high-precision shear measurement is not only a core challenge in weak lensing research but also the key to unlocking the nature of dark matter and cosmological parameters \citep{PRAT2026508, Pantos2026PDU}.

Current shear estimation methods can be primarily divided into two categories: 1. Shape moment-based methods \citep[e.g.][]{KSB1995, Rhodes2000ApJ, Hirata2003}, which calculate the second-order moments of galaxy images to represent their shapes; 2. Model-fitting techniques \citep[e.g.][]{Massey2005MNRAS, Nakajima2007AJ, Miller2007, Zuntz2013}, which fit galaxy images by assuming luminosity distribution models and PSF models, but they have limited adaptability to irregular galaxies or complex morphologies. In recent years, machine learning methods \citep[e.g.][]{Springer2020MNRAS, FORKLENS2024} have shown potential for handling complex data. However, they face challenges such as high dependence on training datasets and insufficient physical interpretability.

The Fourier\_Quad (FQ) method proposed by \cite{Zhang2008MNRAS, Zhang2015JCAP} is a moment-based method. It constructs shear estimators based on the quadrupole and hexadecapole moments of the galaxy power spectrum. 
The effects from the PSF and noise are all corrected in Fourier space in a model-independent way. Its accuracy is in principle unaffected by the signal-to-noise ratio (SNR) and the size of the source image. 
To further approach the theoretical limit of statistical errors, \cite{Zhang2017} introduced the PDF-SYM method, which optimizes the statistical uncertainties of the measurement by symmetrizing the PDF of the shear estimators. Its robustness has been demonstrated in weak lensing measurements with different imaging data sets \citep{Fong2022, Wang2022, Wang2023, Liu2023, Shen2024, Liu2025ApJ, Liu2025ApJsersic}. However, PDF-SYM requires iterative evaluations of the level of PDF symmetry, and the algorithm for higher precision leads to growth in computational complexity, making it challenging to handle the massive data from future surveys like the Large Synoptic Survey Telescope\footnote{www.lsst.org} and EUCLID\footnote{www.euclid-ec.org}. 

The main purpose of this paper is to simplify the shear estimators of FQ so that instead of PDF-SYM, one can use the usual weighted-sum method to evaluate the lensing statistics. The price to pay is likely an increase of the shear bias, and also possible enhancement of statistical uncertainty. In particular, \cite{Zhang2011} argue out that there is no unbiased conventional shear estimators (i.e., averaging over a single quantity per galaxy to estimate each shear component). Therefore, we expect that the new way of the shear estimation should require careful calibrations.

Currently, it is a common practice to calibrate shear biases via realistic image simulations~\citep{Liping2018MNRAS,Mandelbaum2018MNRAS,MacCrann2022MNRAS}.  For example, the HSC first-year analysis generates a suite of mock images matching its observations and derived empirical corrections that reduce the overall shear bias to $\lesssim$ 1\% \citep{Mandelbaum2018MNRAS}. An alternative “self‑calibration” approach is Metacalibration \citep{Metacalibration}, in which the actual survey images are artificially sheared and the shapes re-measured to infer the shear response. It does not rely on external simulation inputs and has been successfully applied to obsevational data~\citep{Gatti2021MNRAS}. There are also some methods that use machine learning to measure shear bias~\citep{Tewes2019AA, Pujol2020AA}.

In this work, we rely on the field-distortion (FD) test proposed by \cite{Zhang2019ApJ} to detect multiplicative and additive shear biases. FD refers to the distortion pattern across the CCD focal plane induced by deviations from the ideal sky-to-focal-plane projection of the optical system. It may manifest as twisting, stretching, or compressing of the image. This calibration method leverages the intrinsic properties of the data itself, without the need for simulations or external datasets. In FD test, we measure the shear values for galaxies at fixed FD values. These galaxies are distributed randomly on the sky, so their average cosmic shear should be zero. By comparing the measured FD signals $g_{\mathrm{f},i}^{\rm obs}$ with the true signals $g_{\mathrm{f},i}$ provided in the catalog, we can determine the multiplicative bias $m_i$ and additive bias $c_i$ for two shear components respectively: 
\begin{equation}\label{eq:gf}
g_{\mathrm{f},i}^{\rm obs} = (1 + m_i) \, g_{\mathrm{f},i} + c_i.
\end{equation}

The structure of this paper is as follows. In \S\ref{sec:FQ}, we introduce the methodology, including the FQ method, the PDF-SYM method, the self-weighted shear estimator (SWSE), the ESD measurement, and the corresponding calibration procedure. In \S\ref{sec:siml}, we test the performance of the method using simulated data. In \S\ref{sec:FD}, we further validate the method with real observations from the DECaLS and HSC surveys. Finally, \S\ref{sec:summary} summarizes our results.

\section{Methodology}\label{sec:FQ}
FQ shear estimators are constructed from the quadrupole moments of the galaxy power spectrum, supplemented by several hexadecapole moments to correct for PSF effects. In the FQ method, each galaxy has five shear estimators: $G_1$, $G_2$, $N$, $U$ and $V$, which are defined as follows:
\begin{eqnarray}\label{eq:gnu}
&&G_1=-\frac{1}{2} \int d^2 \vec{k}\left(k_x^2-k_y^2\right) T(\vec{k}) M(\vec{k})\\ 
&&G_2=-\int d^2 \vec{k} k_x k_y T(\vec{k}) M(\vec{k})\\ 
&&N=\int d^2 \vec{k}\left(k^2-\frac{\beta^2}{2} k^4\right) T(\vec{k}) M(\vec{k})\\ 
&&U=-\frac{\beta^2}{2} \int d^2 \vec{k}\left(k_x^4-6 k_x^2 k_y^2+k_y^4\right) T(\vec{k}) M(\vec{k})\\ 
&&V=-2 \beta^2 \int d^2 \vec{k}\left(k_x^3 k_y-k_x k_y^3\right) T(\vec{k}) M(\vec{k})
\end{eqnarray}
Here, $G_i$ is analogous to the ellipticity component $e_i$, and $N$ is a normalization factor for $G_i$. $U$ is an additional correction term, and $V$ is for calculating $U$ under coordinate rotation, as $(U,V)$ form a pair of spin-4 quantities. $M(\vec{k})$ represents the power spectrum of the galaxy image after background subtraction and Poisson noise removal, and the factor $T(\vec{k})$ is used to transform the PSF into the desired isotropic Gaussian form. $\beta$ is the scale radius of the Gaussian function (for more details, see \cite{Zhang2015JCAP}). 

To obtain an unbiased shear estimate, they propose an unconventional averaging approach to compute the shear:
\begin{equation}
\frac{\left\langle G_i\right\rangle}{\langle N\rangle}=g_i+O\left\langle g^3\right\rangle
\label{ave}
\end{equation}
where $g_i$ is the corresponding reduced shear component and $g_i=\gamma_i/(1-\kappa)$. However, since the shear estimator is not luminosity-normalized, the shear results are dominated by bright galaxies, while the majority of galaxies are faint. To address these issues, \cite{Zhang2017} propose a novel statistical method, the PDF-SYM method. However, considering the convenience of measuring shear, in this work, we propose a weighted averaging shear estimation method, the SWSE method. Next, we will introduce these two statistical methods.

\subsection{PDF-SYM method}\label{sec:pdf-sym}
PDF-SYM is an unconventional method for evaluating various weak lensing statistics. It leverages the full information contained in the PDF of shear estimators  in the cases of shear stacking or shear-shear correlations, as demonstrated in \cite{Zhang2017}. The shear statistics are evaluated by finding the optimal values that symmetrizes the PDF. It has been shown that the PDF-SYM method can bring the statistical errors close to theoretical limits.

The relationship between $G_i$ and the reduced shear ($g_i = \gamma_i / (1 - \kappa)$) can be approximated as
\begin{equation}
\label{eq:GiGs}
G_i=G^{{\rm S}}_i+g_i B_i
\end{equation}
where $G^{{\rm S}}_i$ can be regard as unlensed $G_i$, and
\begin{equation}
  B_i= \left \{
  \begin{array}{rcl}
    N+U & & (i=1)\\
    N-U & & (i=2)\\
  \end{array} \right .
\end{equation}
Based on Eq.\ref{eq:GiGs}, the proposal of PDF-SYM is to observe the symmetric properties of the distribution of 
\begin{equation}
\label{Ghat}
\hat{G}_i=G_i-\hat{g}_iB_i=G^{\rm S}_i+(g_i-\hat{g}_i)B_i.
\end{equation}
where $\hat{g}_i$ is the pseudo-shear value we guess in the symmetrization process. It is demonstrated by \cite{Zhang2017} that, if $\hat{g}_i=g_i$, the PDF of $\hat{G}_i$, $P(\hat{G}_i)$, will be symmetric about 0. Therefore, the shear value can be estimated by finding the $\hat{g}_i$ that best symmetrizes $P(\hat{G}_i)$ relative to zero. Technically, this is achieved by placing bins symmetrically on both side of zero. The number of galaxies in each bin is labeled as $n^{(p)}$, where $p$ is the index of the bins. By varying the values of $\hat{g}_i$, we have
\begin{equation}
\label{chi2g}
\chi^2 (\hat{g_i}) = \frac{1}{2} \sum_{p > 0}\frac{(n^{(p)}-n^{(-p)})^2}{n^{(p)}+n^{(-p)}}
\end{equation} 
as a function of $\hat{g}_i$ to characterize the symmetry of the PDF. Note that the negative bin indices ($-p$) refer to the bins on the negative side of zero. More details can be found in section 3.2 of \cite{Zhang2017}.

However, since the PDF-SYM method requires finding pseudo-shear value that symmetrize the PDF of the shear estimators, each guess of the shear necessitates a calculation of the PDF's symmetry by using Eq. \ref{chi2g}. If the value is guessed for $N$ times, the computational cost is approximately $N$ times that of a weighted averaging method. Additionally, if more bins are used for the PDF, the memory required for computations increases, further decreasing computational speed. As the main purpose of this paper, we introduce a self-weighted averaging method to expedite the shear measurement in practice, and rely on the FD test to guarantee its accuracy.

\subsection{Self-Weighted Shear Estimation}\label{sec:SWSE}
We propose to use the shear estimator itself as the weight for the weighted average estimation, which we call Self-Weighted Shear Estimation (SWSE). For now, we define $e_i=G_i/N$, which is intrinsically biased due to the underlying noise bias and its non-linear nature. The weight $w(|e|)$ of SWSE is dependent on the shear estimators, where $|e|=\sqrt{e_{1}^2+e_{2}^2}$. When we select the weighting function as $w(|e|)=1/|e|$, if $e_i$ is unbiased shear estimators, it can be demonstrated that the bias is 0.5, i.e.,
\begin{equation}\label{meanw}
\frac{\sum_k e_{i}^{(k)}w^{(k)}(|e^{(k)}|)}{\sum_k w^{(k)}}
= \frac{g_i}{2} + \mathcal{O}(g_{1,2}^2).
\end{equation}
The theoretical derivation of this factor is provided in Appendix~\ref{app:derive}. We note that when $e_{i}$ diverges, the product $e_{i} |e|^{-1}$ approaches unity, thereby preventing outliers from driving the final result to divergence. 

\citet{Bernstein2002AJ} proposed that the optimal form of weights that can minimize statistical errors is
\begin{equation}\label{optweight}
w_{\text{opt}}(|e|) \propto 3 - \frac{1 - |e|^2}{|e|} \frac{d \log P(|e|)}{d |e|}.
\end{equation}
$P(|e|)$ denotes the intrinsic ellipticity distribution, which is formally a two-dimensional probability distribution over $(e_1, e_2)$, but is often assumed to be isotropic and hence depends only on $|e|$. Therefore, we also test this weighting scheme in our analysis.

\begin{table} 
\centering  
\caption{Shear estimators and their corresponding weights used in the SWSE scheme.} \label{tab:ei}
\begin{tabular}{l|c|c|c}
\hline
 Scheme & $e_i$ & $w$ & Range of $e_i$  \\
\hline
 SWSE-A & $G_i/N$ & $\mathrm{sgn}(N)/|e|$ & $(-\infty,\infty)$ \\
 [2.4pt]
 SWSE-B & ${\rm Arg}(N +\mathrm{i}\, G_i)$ & $1/|e|$ & $(-\pi,\pi)$ \\
 [2.4pt]
 SWSE-C & $G_i/N$ & $\mathrm{sgn}(N)w_{\rm opt}(|e|)$ & $(-\infty,\infty)$ \\
\hline
\end{tabular}   
\end{table}

We consider three combinations of two weighting schemes and two shear estimators for SWSE method in our analysis, as summarized in Table~\ref{tab:ei}. A factor of $\mathrm{sgn}(N)$ is included in the weights for ``SWSE-A'' and ``SWSE-C'' to maintain sign consistency between the estimator $e_i$ and the signal $G_i$. This treatment prevents unphysical sign reversals in low-SNR regimes ($N < 0$), which we found to be essential for successful FD test to calibrate in real survey data. Alternatively, we consider another shear estimator $e_i={\rm Arg}(N +\mathrm{i}\, G_i)$ in scheme ``SWSE-B''for comparison. It represents the phase angle of the complex number $N + \mathrm{i}G_i$ and is introduced as a complementary shear estimator exhibiting different bias properties, allowing us to assess the performance of the calibration scheme under different shear bias conditions.

For low-SNR samples, the PDF of estimator $e_i = G_i / N$ exhibits pronounced long tails, with a non-negligible fraction of galaxies having very large $|e_i|$ values. These outliers arise from cases where $N$ approaches zero, leading to unstable ratios. If we simply calculate the average without weighting, the result will be highly unstable, with extremely large errors. 
The self-weighting scheme already suppresses the influence of extreme $e_i$ outliers, and therefore no explicit cutoff is required in the present analysis. In practice, however, a loose cutoff (e.g. $|e_i|<1000$) may still be introduced as a numerical safeguard against extremely rare cases with $N\rightarrow0$. Such a cutoff has a negligible impact on our conclusions.

\subsection{Measuring Galaxy-galaxy Lensing}
Galaxy–galaxy lensing describes the correlation between the positions of galaxies or halos and the tangential shear of background sources. It provides a direct probe of the projected mass distribution around galaxies and dark matter halos, independent of their dynamical state or baryonic content. It is widely used to measure halo properties and also serves as a valuable cosmological probe. In this work, we use galaxy–galaxy lensing as a test case to evaluate the performance of different shear estimation methods and the effectiveness of the FD test.

The observable quantity is the excess surface density (ESD), defined as
\begin{equation}\label{eq:esd}
\Delta \Sigma (R) \equiv \overline{\Sigma}(<R) - \Sigma(R) = \Sigma_c\gamma_{t}(R).
\end{equation}
$\overline{\Sigma}(<R)$ represents the average surface density within a projected radius of $R$. The comoving critical surface density is
\begin{equation}\label{sigmac}
\Sigma_c = \frac{c^2 D_{\rm s}}{4\pi G D_{\rm l} D_{\rm ls} (1+z_{\rm l})^2},
\end{equation}
with c the speed of light, $G$ the gravitational constant, and $D_{\rm s}$, $D_{\rm l}$, and $D_{\rm ls}$ the angular diameter distances to the source, to the lens, and from the lens to the source, respectively.

For the PDF-SYM method, measuring the ESD follows a procedure similar to shear estimation described in Sec.\ref{sec:pdf-sym}, with the only modification being the inclusion of the critical surface density factor $\Sigma_c$. Specifically, Eq.~\ref{Ghat} is rewritten as
\begin{equation}
\Sigma_c \hat{G}_t = \Sigma_c G_t - \widetilde{\Delta\Sigma}(N + U_t),
\end{equation}
where $G_t$ is a spin-2 quantity and $U_t$ is a spin-4 quantity. The parameter $\widetilde{\Delta\Sigma}$ denotes a pseudo-ESD value introduced to symmetrize the PDF of $\Sigma_c \hat{G}_t$. The value of $\widetilde{\Delta\Sigma}$ that maximizes the symmetry of the PDF corresponds to the measured ESD.

For the SWSE method, in addition to the shear-estimation weights defined in Table~\ref{tab:ei}, we further incorporate an inverse-variance factor to obtain a minimum-variance estimator of $\Delta\Sigma$. Since the variance of an individual ESD estimate scales as $\Sigma_c^2$, the optimal weight for each source galaxy is given by the product of the shear-estimation weight and an additional factor of $1/\Sigma_c^2$. This weighting scheme naturally downweights source galaxies with large $\Sigma_c$, which contribute disproportionately higher noise, thereby ensuring a minimum-variance ESD measurement. The ESD estimator is therefore written as
\begin{equation}
\widehat{\Delta\Sigma}
= \frac{\sum_k \Sigma_c e_{t}^{(k)} w^{\prime(k)}}{\sum_k w^{\prime(k)}},
\end{equation}
with
\begin{equation}
w^{\prime} = \frac{w}{\Sigma_c^2}.
\end{equation}
The modified weight $w^{\prime}$ is also adopted in the FD tests to ensure consistency between shear-bias measurements and ESD estimation. In Appendix~\ref{app:weights}, we show that the weights used in the FD test should be kept consistent with those adopted in the ESD measurement in order to properly correct for shear bias.

\subsection{Calibration for the shear estimators}

It is important to note that the FD contribution and additive biases have already been removed from the shear estimators in the catalog used in our analysis \citep{2022arXiv220602434Z, Liu2024zlh}. Therefore, before applying our calibration procedure, these contributions should be added back to recover the original estimators:
\begin{equation}
G_{\mathrm{f,1}}=G_1+(g_{\mathrm{f,1}}+c_1^\prime)(N+U)+(g_{\mathrm{f,2}}+c_2^\prime)V,
\end{equation}
\begin{equation}
G_{\mathrm{f,2}}=G_2+(g_{\mathrm{f,2}}+c_2^\prime)(N-U)+(g_{\mathrm{f,1}}+c_1^\prime)V,
\end{equation}
where $g_{\mathrm{f},i}$ is the FD values provided in catalog, and $c_i^\prime$ represents the additive shear bias measured from the full galaxy sample. 

We apply the multiplicative and additive bias corrections measured from the FD test separately to each shear estimator. For the PDF-SYM method, the corrected quantities are
\begin{equation}\label{eq:pdf_g1corr}
G_1^{\rm corr}=\frac{G_{\mathrm{f,1}}-c_1 (N+U)}{1+m_1}-g_{\mathrm{f,1}}(N+U)-g_{\mathrm{f,2}}V,
\end{equation}
\begin{equation}\label{eq:pdf_g2corr}
G_2^{\rm corr}=\frac{G_{\mathrm{f,2}}-c_2 (N-U)}{1+m_2}-g_{\mathrm{f,2}}(N-U)-g_{\mathrm{f,1}}V.
\end{equation}
For the conventional shear estimators, the calibration is applied as
\begin{equation}\label{eq:swse_gcorr}
e_i^{\rm corr}=\frac{e_{\mathrm{f},i} - c_i}{1+m_i}-g_{\mathrm{f},i},
\end{equation}
where $e_{\mathrm{f},i}$ denotes either $G_{\mathrm{f},i}/N$ or ${\rm Arg}(N + \mathrm{i}\,G_{\mathrm{f},i})$ as used in this work. We should be aware that the weights during the correction process should remain in their uncorrected state.

Note that the above procedures are essential for removing the residual additive biases from FD. This is however not easily achievable in other shear measurement methods, as the FD signals are typically believed to be removed once for all from the shear estimators.

\section{Results from Mock Galaxies}\label{sec:siml}
In this section, we generate realistic galaxy images to evaluate the calibration performance of different shear estimation methods. In our simulations, we input both the Cartesian shear components $g_{\mathrm{f,1}}$ and $g_{\mathrm{f,2}}$, ranging from $-0.04$ to $0.04$, as well as the tangential shear $g_t$ corresponding to the ESD signals. By first measuring the shear biases of various estimators for $g_{\mathrm{f,1}}$ and $g_{\mathrm{f,2}}$ fields, we can then apply the resulting calibration to the tangential shear $g_t$. This procedure captures the core principle of our FD test.

\subsection{Mock Galaxies}
We adopt galaxy properties from the simulated galaxy catalogue of \cite{Shunsheng}, which was developed for the KiDS-Legacy weak-lensing image simulations. In this catalogue, the large-scale galaxy distribution and photometric properties are based on the SURFS cosmological $N$-body simulations \citep{Elahi2018MNRAS} combined with the Shark semi-analytic galaxy formation model \citep{Lagos2018MNRAS}, while the galaxy morphological parameters are learned from HST observations of the COSMOS field. For more information, please refer to \cite{Shunsheng}.

Each galaxy is modelled using a Sérsic surface-brightness profile. The morphological parameters of the Sérsic model include the half-light radius, axis ratio, and Sérsic index. These parameters provide realistic distributions of galaxy sizes and intrinsic ellipticities. The orientations of galaxies are randomly assigned. The simulated galaxy catalogue spans a wide redshift range up to $z_s \sim 2$, ensuring a realistic distribution of background sources for weak-lensing analyses. The magnitude and size distributions of galaxies are also consistent with those observed in the KiDS survey. Using these parameters, we generate galaxy images and convolve them with a Gaussian PSF of $\sigma = 1.6$ grids. Uniform Gaussian background noise is then added to mimic observational conditions.

For the tangential shear mentioned above, each background galaxy ($z_s$) is assigned a projected separation from a foreground lens ($z_l=0.1$), with the number of galaxies at different radii distributed proportionally to the area of the corresponding annulus. The tangential shear $g_t$ is then computed from the input ESD profile together with the corresponding $\Sigma_c(z_l,z_s)$, and applied according to the lens–source geometry.

\subsection{Shear Measurements and Biases}

\begin{figure*}[ht!]
\centering
\includegraphics[width=0.9\textwidth]{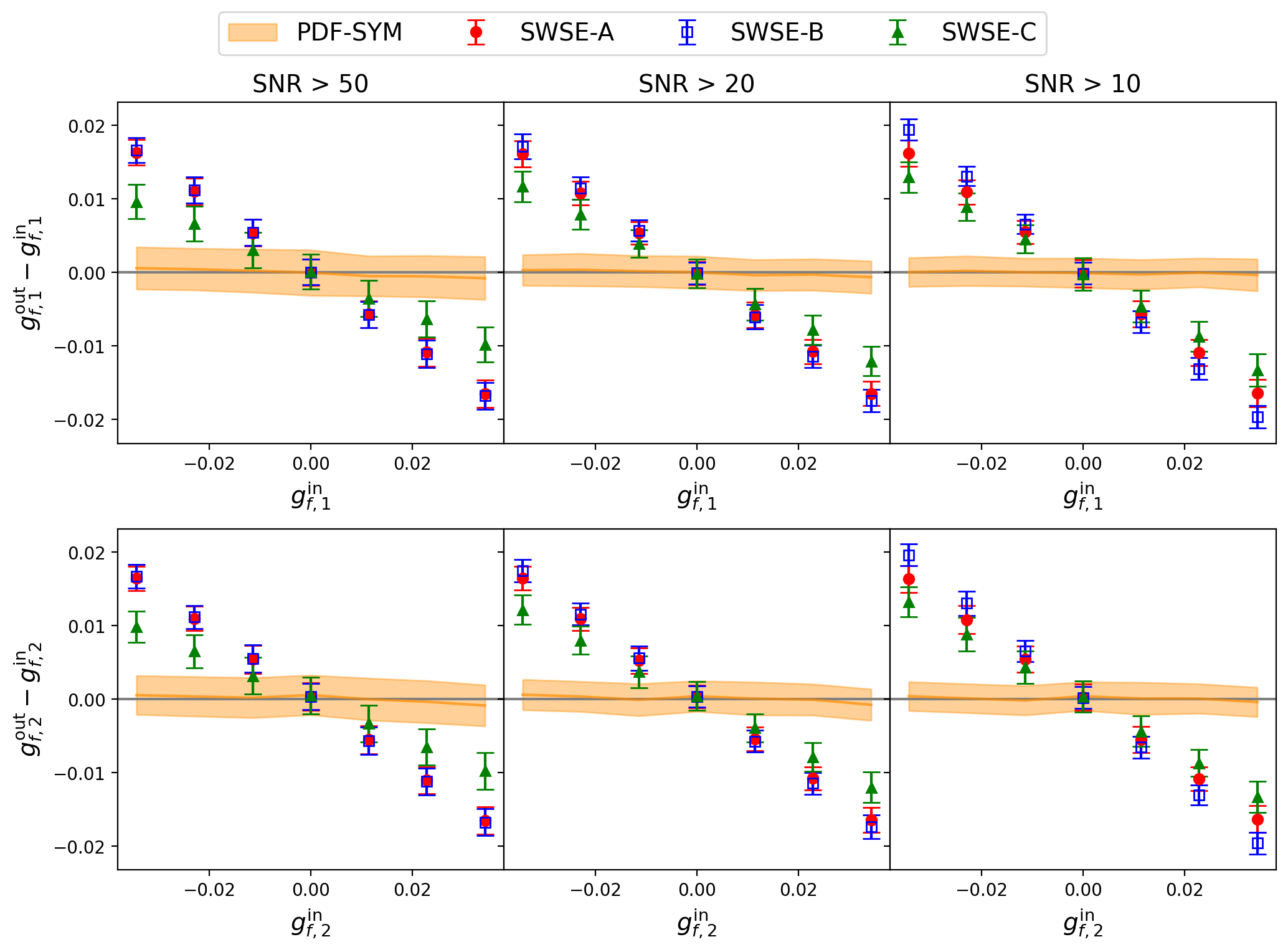}
\caption{Comparison of shear measurement biases for the two shear components, $g_{f,1}$ (top panels) and $g_{f,2}$ (bottom panels), in mock data. The vertical axis shows the difference between the measured and input shear, $g_{\mathrm{f},i}^{\mathrm{out}} - g_{\mathrm{f},i}^{\mathrm{in}}$. Each column corresponds to a different SNR cut. The orange shaded regions represent the results from the PDF-SYM method. The red solid points and blue open points refer the results using SWSE method with shear estimators $e_i=G_i/N$ and $e_i={\rm Arg}(N+G_i\,{\rm i})$ respectively. \label{fig:g1g2}}
\end{figure*}

\begin{figure}[ht!]
\centering
\includegraphics[width=0.4\textwidth]{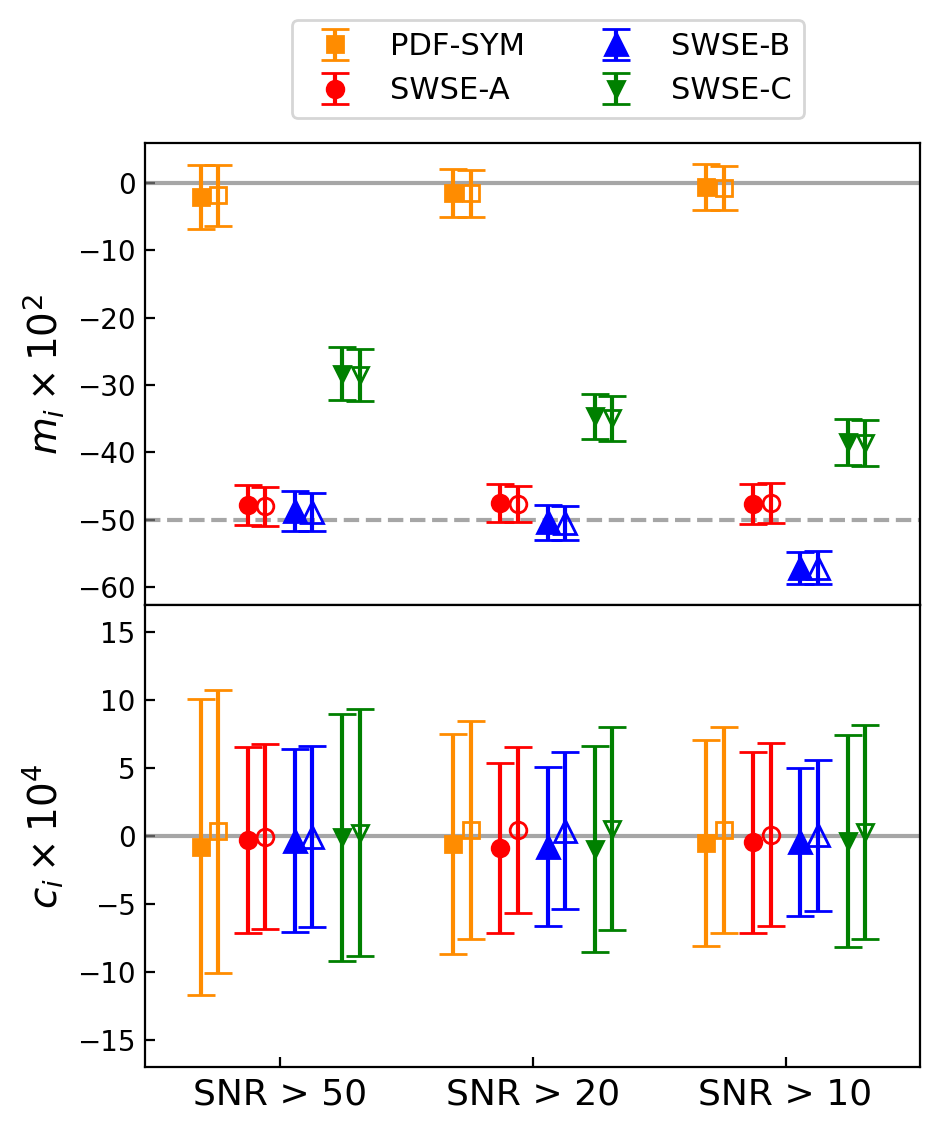}
\caption{The multiplicative and additive biases of $g_{\rm f,1}$ (solid markers) and $g_{\rm f,2}$ (open markers) for both the PDF-SYM and SWSE methods with shear estimators $e_i=G_i/N$ (red) and $e_i={\rm Arg}(N+G_i\,{\rm i})$ (blue) under different SNR cuts in mock data.
\label{fig:mc}}
\end{figure}

Figure~\ref{fig:g1g2} presents the recovered shear signals for $g_{\mathrm{f,1}}$ and $g_{\mathrm{f,2}}$ in the mock simulations, obtained using the PDF-SYM and SWSE methods under different SNR cuts. Figure~\ref{fig:mc} shows the corresponding shear biases for $g_{\mathrm{f,1}}$ (solid symbols) and $g_{\mathrm{f,2}}$ (open symbols). Firstly, we note that SWSE approaches effectively suppresses the impact of extreme $e_i$, yielding statistical uncertainties comparable to those of the PDF-SYM method. Regarding shear biases, the PDF-SYM method delivers nearly unbiased shear estimates, with both multiplicative and additive biases consistent with zero. In contrast, the SWSE method exhibits distinct multiplicative bias behaviors depending on the specific combination of weighting schemes and shear estimators. For “SWSE-A”, the multiplicative bias $m_i$ remains remarkably stable across a wide range of SNR, with values close to $-0.5$. This aligns closely with the theoretical prediction from Eq. \ref{meanw}. The slight deviation of $m_i$ above the theoretical value $-0.5$ is attributed to the intrinsic bias of the $G_i/N$ estimator, arising from its non-linear response to noise. For "SWSE-B", the results are nearly identical to those of "SWSE-A" in the high-SNR regime. However, as the SNR decreases, the multiplicative bias increases significantly. For "SWSE-C", the optimal weight is calculated based on the sample with ${\rm SNR}>10$ as follows:
\begin{equation}\label{wopt_mock}
    W_{\mathrm{opt}}^{\mathrm{mock}}(|e|) = 
\begin{cases} 
10^{-11 |e|} & (|e| < 1.2) \\[4pt]
|e|^{-2}/14.5 & (|e| \geq 1.2)
\end{cases}
\end{equation}
We find that the multiplicative bias of this scheme varies with SNR, as expected. Regarding the additive bias, no detectable signal is observed across all schemes, owing to the idealized nature of our simulation data.

\subsection{Calibration for ESD Measurements}

\begin{figure*}[ht!]
\centering
\includegraphics[width=0.8\textwidth]{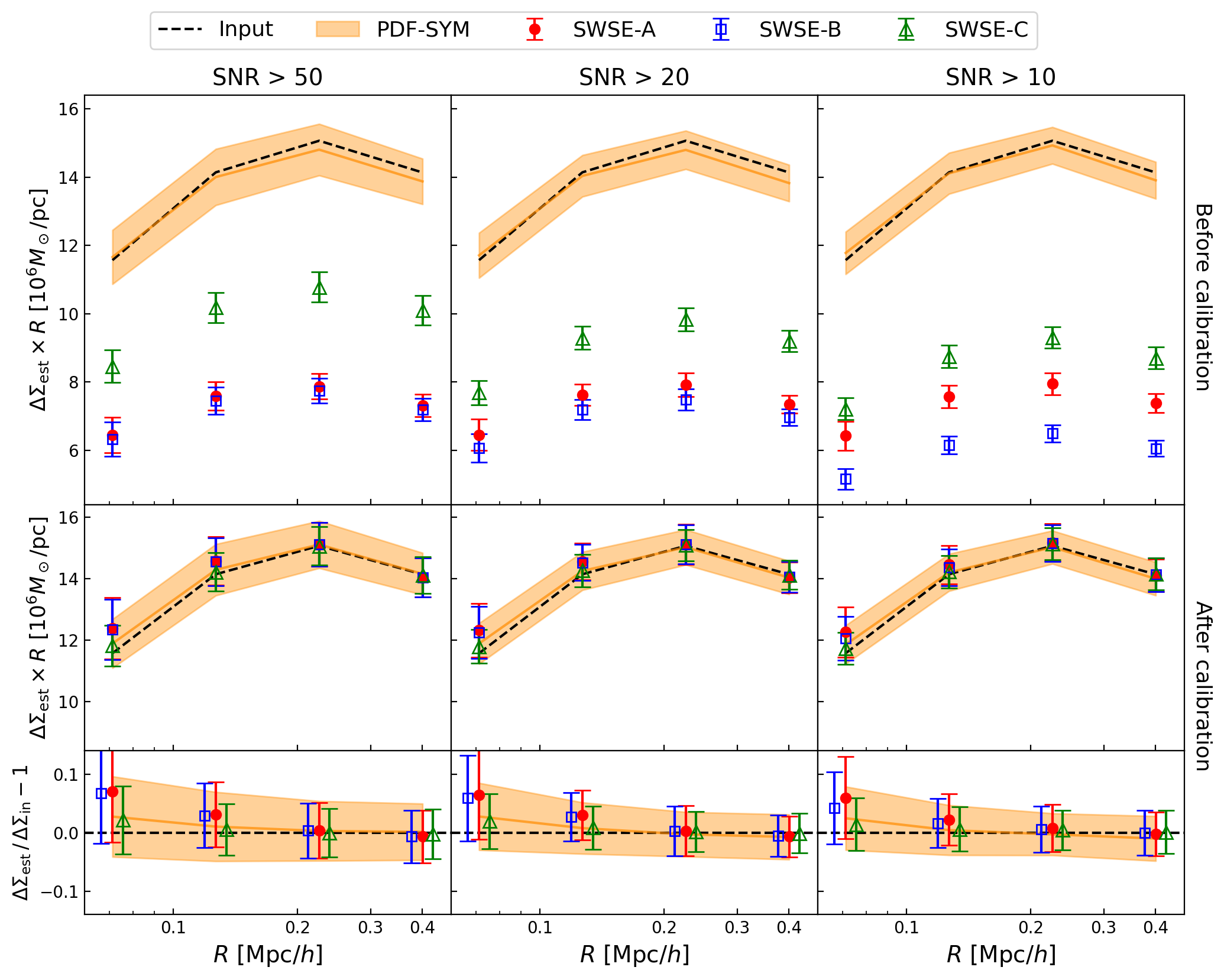}
\caption{ESD signals measured using different shear estimation methods in mock data, before (top panels) and after (bottom panels) applying the multiplicative shear calibration obtain from Fig.\ref{fig:g1g2}. Each column corresponds to a different signal-to-noise ratio (SNR) cut. The orange shaded regions represent the results from the PDF-SYM method. The red solid points and blue open points refer the results using SWSE method with shear estimators $e_i=G_i/N$ and $e_i={\rm Arg}(N+G_i{\rm i})$ respectively. The black dashed lines indicate the input ESD signals.  \label{fig:siml_ESD}}
\end{figure*}

Using the shear biases measured from $g_{\rm f,1}$ and $g_{\rm f,2}$ (Figure \ref{fig:mc}), we apply the corresponding calibration to the simulated ESD signals. For each galaxy, the shear estimator is corrected by its multiplicative bias according to $e_i^{\rm cal}=e_i/(1+m_i)$. Here we correct only for the multiplicative bias, since the additive bias is small in the mock data and it is effectively averaged out during the azimuthal stacking in the ESD measurement. The upper panels of Figure \ref{fig:siml_ESD} display the uncorrected ESD measurements for the three estimations. In agreement with the results from the $g_{\rm f,1}$ and $g_{\rm f,2}$ tests, the ``SWSE-A'' scheme yields results that are systematically lower by about 50\% compared to the PDF-SYM method, and the results of ``SWSE-B'' and ``SWSE-C'' become increasingly underestimated at lower SNR. Upon applying the shear bias calibration (bottom panels), all methods successfully reproduce the input ESD profiles (black dashed lines), exhibiting excellent agreement with the truth. Notably, ``SWSE-C'' yields the results with the smallest statistical uncertainties. These mock tests validate the efficacy of the FD test in calibrating self-weighted shear estimators across various weighting schemes.

\section{ONSITE SHEAR CALIBRATION in real data}\label{sec:FD}
In real observations, shear measurements are affected by a variety of unknown systematics and selection effects beyond those introduced by shear estimator biases, making it essential to validate calibration methods on real data. In this section, we apply our SWSE methods and calibration framework to real survey data to assess their performance and to evaluate the effectiveness.

\subsection{Data}\label{sec:data}
The data includes foreground galaxy cluster catalogs and two background shear catalogs from different surveys. For the background shear catalogs, we use two shear catalogs for multiple validations. One is from the Dark Energy Camera Legacy Survey (DECaLS) and the other one is from the third public data release of the Hyper Suprime-Cam (HSC) Survey. Both datasets are processed using the FQ shear measurement pipeline \citep{2022arXiv220602434Z, Liu2024zlh}. According to the FD tests presented in these works, the multiplicative and additive biases for the sample used in this study are consistent with zero within the statistical uncertainties. Therefore, no additional bias correction is applied to the shear catalog. In the galaxy-galaxy lensing analysis, we ensure that background galaxies have a photo-$z$ larger than the lens redshift by 0.2, i.e., $z_s > z_l + 0.2$, to reduce the dilution of cluster and background galaxies. 

Our galaxy cluster catalog comes from \cite{yang2021extended}, based on the Dark Energy Spectroscopic Instrument (DESI; \cite{Dey2019AJ}) Image Legacy Surveys DR9. In this study, we select galaxy clusters with halo masses in the range of $10^{14}-10^{14.5} M_\odot/h$ for stronger shear signals.  

DECaLS is one of the three public imaging surveys within DESI, designed primarily for target selection under relatively poor seeing conditions (typically $\sim 1.5^{\prime\prime}$). The survey covers roughly 10,000 deg$^2$ in the $g$, $r$, and $z$ bands, with the $z$ band offering the highest image quality. Accordingly, in our tests we use only the $z$-band data and exclude galaxies with $\mathrm{SNR} < 20$. We further find that galaxies fainter than magnitude 21 exhibit large photometric-redshift uncertainties, which lead to a systematic suppression of the measured ESD signal, which is an effect that cannot be recovered through shear calibration alone. Therefore, we restrict our DECaLS source sample to galaxies with $z$-band magnitude $< 21$. The optimal weight for the ``SWSE-C'' scheme for this sample is given by
\begin{equation}\label{wopt_decals}
    W_{\mathrm{opt}}^{\mathrm{DECaLS}}(|e|) = 
\begin{cases} 
10^{1.2-1.2 |e|} & (|e| < 1) \\[4pt]
|e|^{-3} & (|e| \geq 1)
\end{cases}
\end{equation}

HSC provides exceptionally deep optical imaging, reaching a limiting magnitude of $\sim 26$ over approximately 1,400 deg$^2$. The survey includes five bands ($g$, $r$, $i$, $z$, and $y$), with the $i$ band delivering the highest image quality.
In our analysis, we use only the $i$-band images and impose a minimum SNR of 10 on the source galaxies. The optimal weight for the ``SWSE-C'' scheme for HSC sample is given by
\begin{equation}\label{wopt_decals}
    W_{\mathrm{opt}}^{\mathrm{HSC}}(|e|) = 
\begin{cases} 
10^{-1.4 |e|} & (|e| < 0.5) \\[4pt]
|e|^{-2}/20 & (|e| \geq 0.5)
\end{cases}
\end{equation}

\subsection{Results}\label{sec:ggl}

\begin{figure*}[ht!]
\centering
\includegraphics[width=0.8\textwidth]{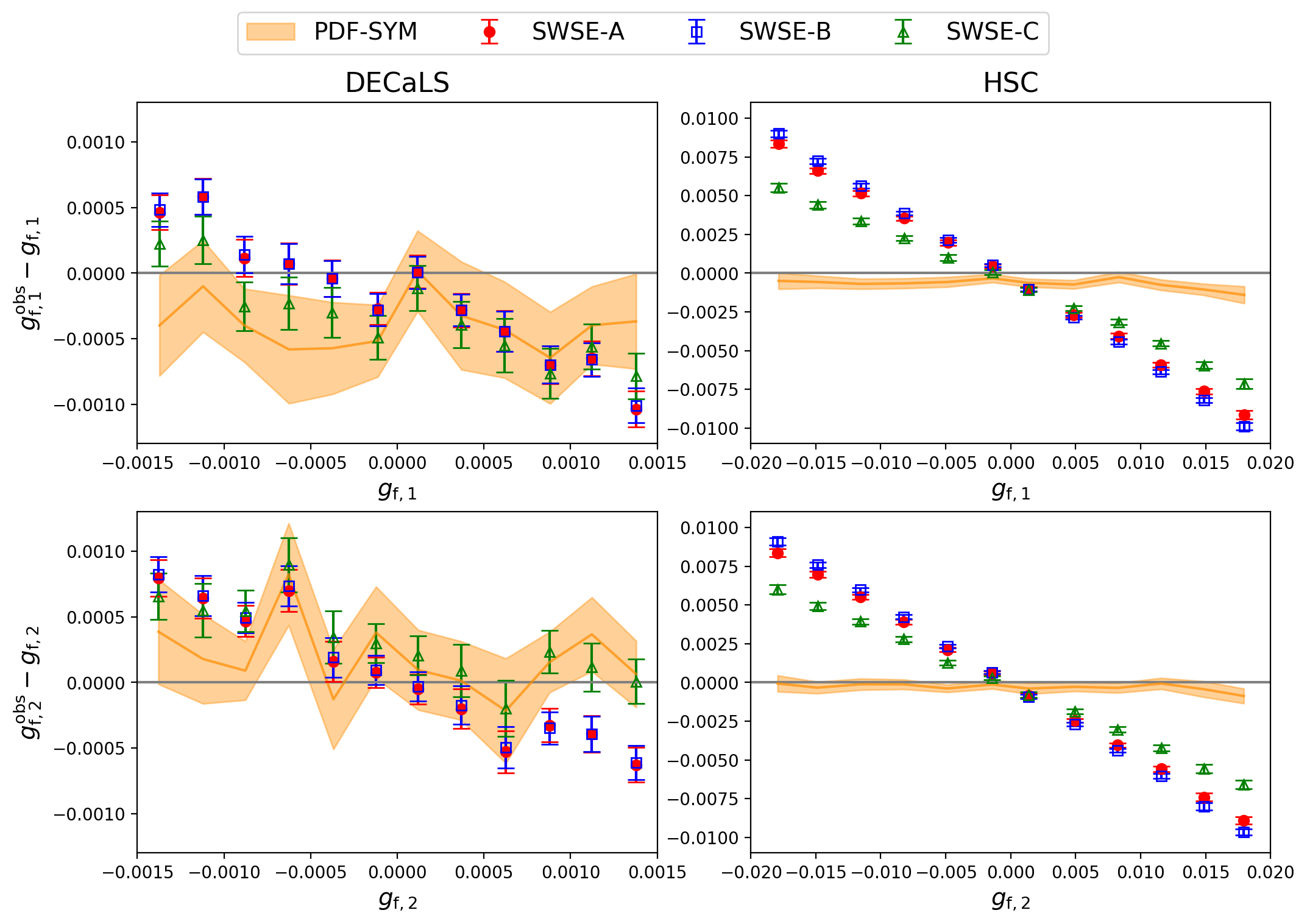}
\caption{Field distortion test for the PDF-SYM method (orange shaded regions) and the SWSE methods. The left and right columns correspond to background galaxies from the DECaLS and HSC surveys, respectively, used in the galaxy–galaxy lensing measurements shown in Fig.~\ref{fig:ggl}.  \label{fig:mc_ggl}}
\end{figure*}

\begin{table*} 
\centering  
\caption{The multiplicative and additive biases of the two shear components of the background galaxies in ESD measurements using the PDF-SYM and SWSE methods. The background galaxy catalogs come from the DECaLS and HSC surveys, respectively. The ESD signals and their calibrations for different subsamples are displayed in Figure \ref{fig:ggl}. The shear biases are derived from the fitting results in Figure \ref{fig:mc_ggl}, using the FD test.} \label{tab:fd-mc}
\begin{tabular}{l|l|cc|cc}
\hline
 Samples & Method & $m_1 \times 10^{2}$ & $c_1 \times 10^{4}$ & $m_2 \times 10^{2}$ & $c_2 \times 10^{4}$  \\
[2.4pt]
\hline
 DECaLS &PDF-SYM &$-3.0 \pm 11.5$ &$-3.9 \pm 1.0$ & $-4.5 \pm 9.8 $ &$1.7 \pm 0.9$ \\
&SWSE-A &$-50.7 \pm 4.6$ &$-1.8 \pm 0.4$ & $-51.7 \pm 4.6 $ &$0.6 \pm 0.4$ \\
  &SWSE-B &$-51.0 \pm 4.5$ &$-1.7 \pm 0.4$ & $-52.6 \pm 4.5 $ &$0.8 \pm 0.4$ \\
  &SWSE-C &$-32.9 \pm 6.0$ &$-3.2 \pm 0.5$ & $-25.7 \pm 6.0 $ &$3.1 \pm 0.5$ \\
 [2.4pt]
\hline
HSC &PDF-SYM &$-1.0 \pm 1.0$ &$-6.3 \pm 1.0$ & $-0.8 \pm 1.0 $ &$-2.8 \pm 1.0$ \\
&SWSE-A &$-47.9 \pm 0.5$ &$-3.4 \pm 0.5$ & $-48.1 \pm 0.5 $ &$-1.3 \pm 0.5$ \\
  &SWSE-B &$-51.9 \pm 0.5$ &$-3.3 \pm 0.4$ & $-52.2 \pm 0.5 $ &$-1.3 \pm 0.4$ \\
  &SWSE-C &$-34.4 \pm 0.6$ &$-5.8 \pm 0.5$ & $-35.0 \pm 0.6 $ &$-2.3 \pm 0.5$ \\
 [2.4pt]
\hline
\end{tabular}  
\end{table*}

We employ the same background galaxy samples in FD test as used in the ESD measurements. The results for various methods and datasets are presented in Figure \ref{fig:mc_ggl}, where the x-axis represents the real FD values ($g_{\mathrm{f,1}}$ and $g_{\mathrm{f,2}}$) provided in the catalog, and the y-axis shows the residuals between our measurements and the truth. Table \ref{tab:fd-mc} summarizes the multiplicative and additive biases obtained by fitting the linear model defined in Eq.~\ref{eq:gf}. We observe a slight inconsistency between the two multiplicative bias components ($m_1$ and $m_2$) for ``SWSE-C'' in the DECaLS data, whereas other methods yield more consistent results. This discrepancy likely stems from certain systematics coupled with the PDF of the shear estimators. In real survey data, the observed biases originate not only from the intrinsic properties of the weights and estimators identified in simulations but also from measurement-induced systematics within the data processing pipeline. Notably, non-zero additive biases are detected across all datasets.

\begin{figure*}[ht!]
\centering
\includegraphics[width=0.8\textwidth]{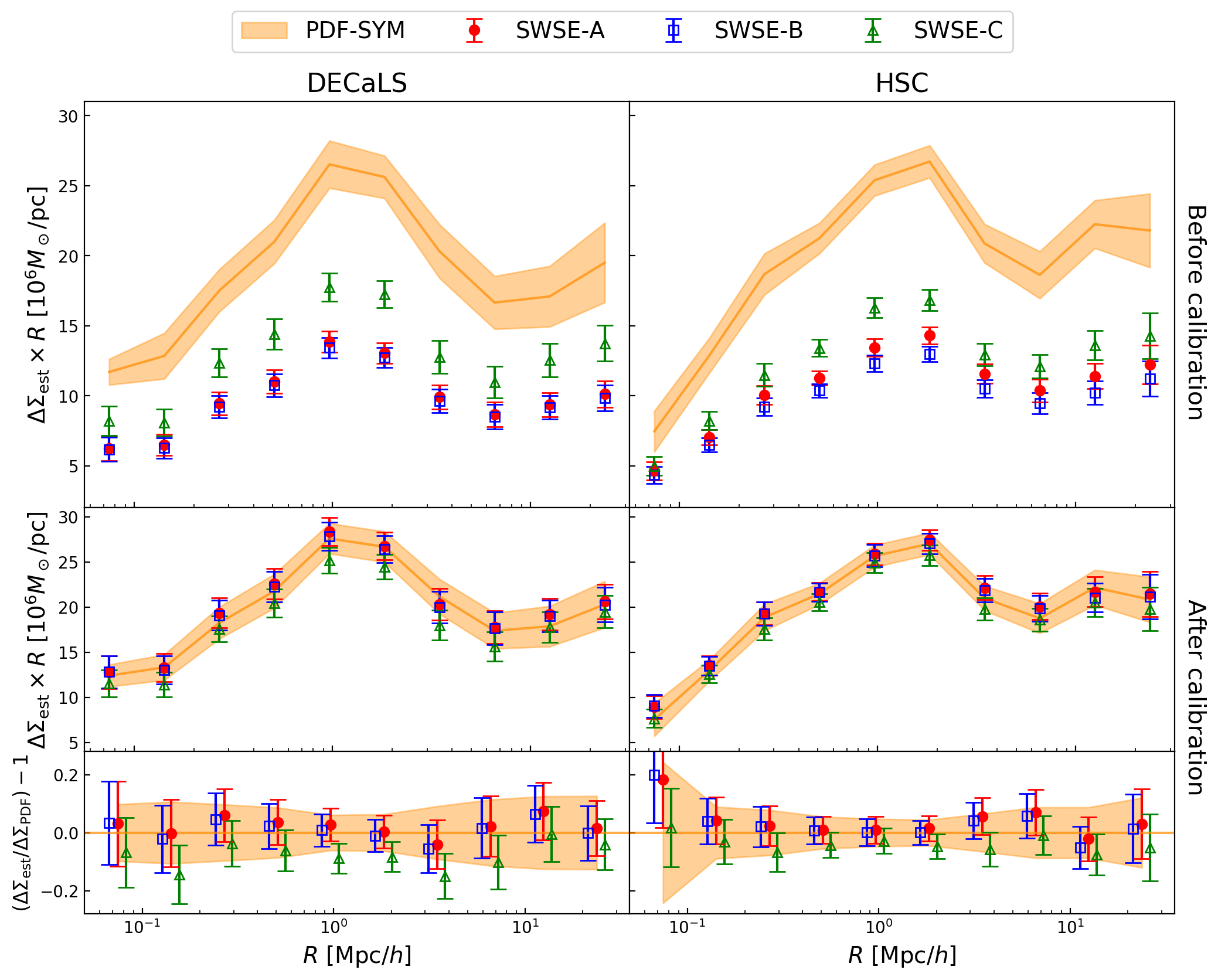}
\caption{ESD measurements derived using the PDF-SYM method (orange shaded regions) and the SWSE methods. The left and right columns correspond to background galaxy samples from the DECaLS and HSC surveys, respectively. The upper panels show the results without shear calibration, while the lower panels present the measurements after correcting for shear biases. \label{fig:ggl}}
\end{figure*}

Figure \ref{fig:ggl} presents the ESD measurements for the DECaLS (left) and HSC (right) surveys. The upper panels show the raw results obtained using the PDF-SYM method and various SWSE schemes before applying any shear calibration. The lower panels display the corresponding calibrated results, along with the fractional difference between the SWSE-based ESD and the PDF-SYM benchmark. Our analysis reveals that the calibrated results from both ``SWSE-A'' ($e_i = G_i/N$) and ``SWSE-B'' ($e_i = {\rm Arg}(N + \mathrm{i}G_i)$) are in excellent agreement with the PDF-SYM reference, demonstrating the reliability of the FD test in mitigating the intrinsic biases of SWSE in observations. However, ``SWSE-C'' exhibits a slight but noticeable systematic deficit. We hypothesize that this behavior originates from the coupling between the optimal weighting scheme and certain focal-plane-dependent systematics, a topic we will discuss in \S\ref{sec:summary}. Regardless of this minor discrepancy, our results validate the overall efficacy of the FD test for SWSE in large-scale galaxy surveys.

\section{Conclusion and Discussion}\label{sec:summary}
This work introduces a shear estimation method, named SWSE, which adopts a self-weighting scheme that naturally suppresses shape noise and balances the contributions from bright and faint galaxy populations, thereby improving the stability of shear estimation. Since SWSE is intrinsically biased, we further investigate whether FD tests can robustly diagnose and calibrate the associated multiplicative and additive shear biases under different estimators and weighting schemes. Using both simulated data and real survey observations, we demonstrate that FD test reliably calibrates these biases for different estimators and weighting schemes. After calibration, SWSE achieves statistical precision and accuracy comparable to those of PDF-SYM while being substantially more computationally efficient. In this work, the performance of SWSE and FD test is primarily validated using ESD measurements, while the application to shear-shear correlation statistics will be presented in a companion paper (Liu et. al 2026), where similarly promising performance is obtained. These results establish the combination of SWSE and FD test as a robust framework for future large weak lensing surveys in the era of increasingly massive datasets.

\begin{figure*}[ht!]
\centering
\includegraphics[width=0.8\textwidth]{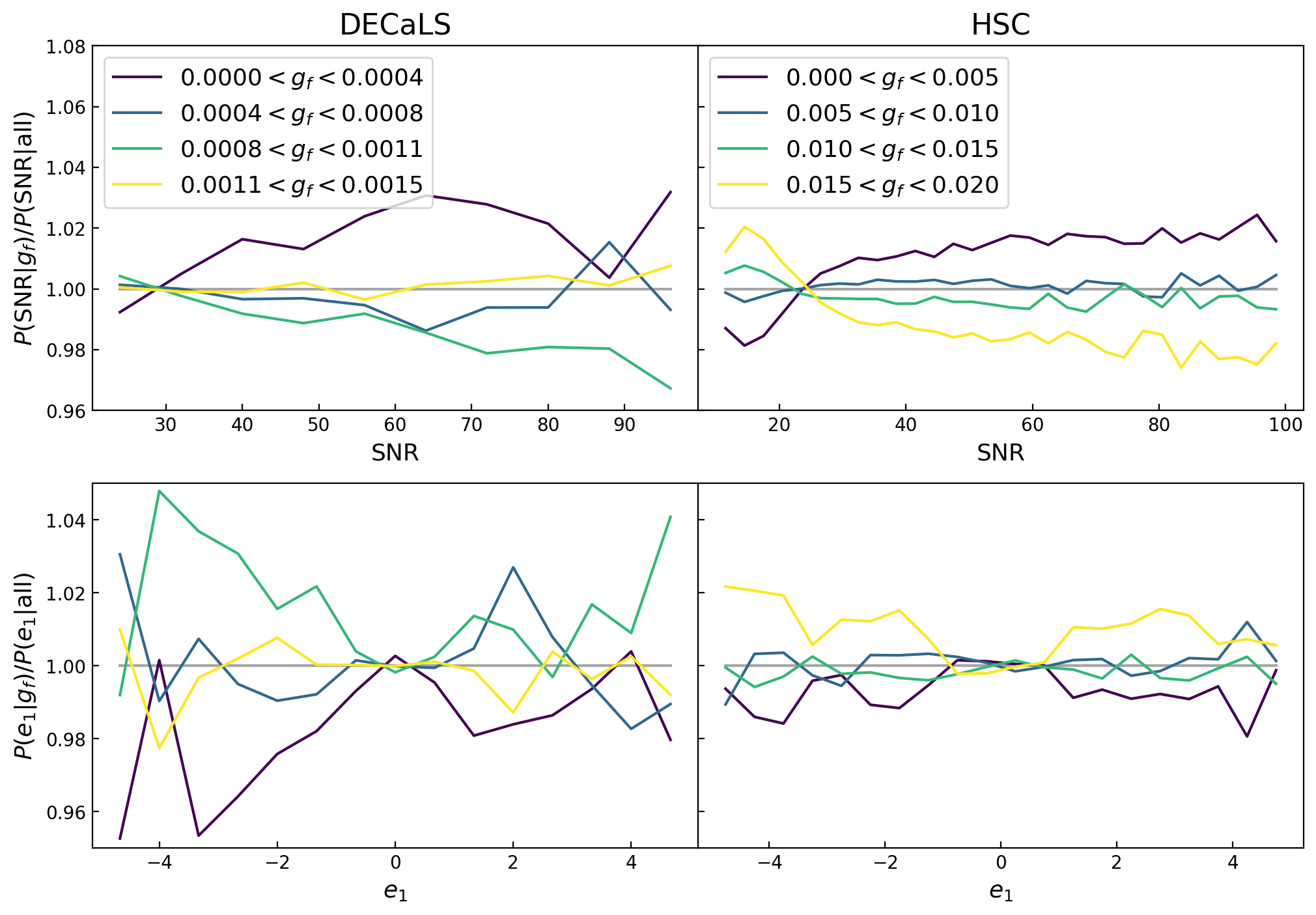}
\caption{Distributions of galaxy SNR (top panels) and the estimator $e_1 = G_1/N$ (bottom panels) in different $|g_f|$ bins for DECaLS (left) and HSC (right). In each case, the distributions are normalized and then divided by the corresponding normalized distribution of the full sample, such that deviations from unity indicate differences relative to the overall population. \label{fig:pdf_gfsnr}}
\end{figure*}

While the FD test has been demonstrated to accurately recover shear biases under idealized conditions (Sec.~\ref{sec:siml}), its efficacy in practical applications can be compromised by instrumental systematics that are spatially correlated with the FD. Specifically, if image properties exhibit non-uniform distributions across the focal plane, the resulting coupling between local noise and the FD signal may introduce biased calibration. Figure~\ref{fig:pdf_gfsnr} presents the distributions of galaxy SNR (top panels) and the estimator $e_1 = G_1/N$ (bottom panels) across different $|g_{\rm f}|$ bins, normalized by their overall distributions for DECaLS and HSC. It is evident that in both surveys, galaxy SNR is slightly coupled with $|g_{\rm f}|$, generally showing higher SNRs near the focal-plane center. This behavior can be attributed to spatial variations in the PSF, optical aberrations, and effective depth across the field of view, which collectively yield higher SNRs and more reliable shape measurements toward the focal-plane center. Hence, populations with lower SNRs exhibit significantly broader and more extended distributions $P(e_i)$ as shown in the lower panel of Figure~\ref{fig:pdf_gfsnr}.

However, such spatially varying image properties imply that the shear bias itself may also vary across the focal plane. In this situation, the underlying assumption of a linear response between the measured shear bias and the FD signal is violated, as the bias dependent on additional observational variables, such as SNR, PSF properties, and focal-plane position. Consequently, a simple linear fit to the FD test does not necessarily recover the true mean shear bias of the sample, and the resulting calibration may therefore be inaccurate. However, our mock results indicate that the biases of the PDF-SYM method and the SWSE-A scheme remain largely insensitive to the distribution of shear estimators, and hence to SNR variations, as shown in Figure~\ref{fig:mc}. In contrast, the SWSE-C scheme exhibits a clear dependence of shear bias on SNR, implying that its shear bias varies with focal-plane position. This dependence may partially explain the slight underestimation observed in the calibrated results of the ``SWSE-C'' scheme in Figure \ref{fig:ggl}. Extending the FD framework to account for spatially varying shear biases across the focal plane will therefore be an important direction for future work.

\begin{acknowledgements}
This work is supported by the National Key Basic Research and Development Program of China (2023YFA1607800, 2023YFA1607802), the NSFC grant (12573004), and the science research grants from China Manned Space Project (No. CMS-CSST-2021-A01). The computations in this paper were run on the $\pi$ 2.0 cluster supported by the Center of High Performance Computing at Shanghai Jiaotong University. 
\end{acknowledgements}

\appendix                  
\section{The multiplicative bias of SWSE-Inv}\label{app:derive}

SWSE-Inv is a weighted averaging method that uses the inverse of the estimator as the weight, as shown in Eq.~\ref{meanw}. Since the weight depends explicitly on the estimator itself, this weighting scheme generally introduces a multiplicative bias. Importantly, this bias is not specific to shear estimators, but arises generically for any random variable whose PDF is symmetric about a small value. Here, we derive the multiplicative bias in this general case.

Suppose there are two sets of random numbers, $e_{S 1}$ and $e_{S 2}$, which follow same PDFs and symmetric about 0, i.e., $P(e_{S 1})=P(e_{S 2})$ and $\langle e_{S 1} \rangle =\langle e_{S 2}\rangle = 0$. These two sets of random numbers can be considered as the unlensed conventional shear estimators. After respectively adding small shears $g_1$ and $g_2$ to them, we have
\begin{equation}
 e_i=e_{S i}+g_i.
\end{equation}
Additionally, Eq.~\ref{meanw} can also be expressed as
\begin{eqnarray}\label{eq:swse_app}
\langle e_i\rangle_{\rm SWSE-Inv} 
&=&  \frac{\int_{-\infty}^{\infty} {\rm d} e_1 \int_{-\infty}^{\infty} {\rm d} e_2 P\left(e_1, e_2\right) \frac{e_i}{\sqrt{e_1^2+e_2^2}}}{\int_{-\infty}^{\infty} {\rm d} e_1 \int_{-\infty}^{\infty} {\rm d} e_2 P\left(e_1, e_2\right) \frac{1}{\sqrt{e_1^2+e_2^2}}} 
\end{eqnarray}
where $P\left(e_1, e_2\right)$ is the joint distribution of $e_1$ and $e_2$. Since $e_i=e_{Si}+g_i$, we have $P \left(e_1, e_2\right)=P_S \left(e_{S1}, e_{S2}\right)$, where $P_S$ is the distribution of $\left(e_{S1}, e_{S2}\right)$. Since $g_1$ and $g_2$ are tiny, we can apply a Taylor expansion to Eq.\ref{eq:swse_app}, and here we take $g_1$ as example, 
\begin{eqnarray}\label{eq:tylor}
\langle e_1\rangle_{\rm SWSE-Inv} 
&=&  \frac{\int_{-\infty}^{\infty} {\rm d} e_{S1} \int_{-\infty}^{\infty} {\rm d} e_{S2} P_S \left(e_{S1}, e_{S2}\right) [(e_{S1}+g_1)^2+(e_{S2}+g_2)^2]^{-\frac{1}{2}}(e_{S1}+g_1)}{\int_{-\infty}^{\infty} {\rm d} e_{S1} \int_{-\infty}^{\infty} {\rm d} e_{S2} P_S \left(e_{S1}, e_{S2}\right) [(e_{S1}+g_1)^2+(e_{S2}+g_2)^2]^{-\frac{1}{2}}} \nonumber\\
&=& \frac{\int_{-\infty}^{\infty} {\rm d} e_{S1} \int_{-\infty}^{\infty} {\rm d} e_{S2} P_S \left(e_{S1}, e_{S2}\right) [(e_{S1}^2+e_{S2}^2)^{-\frac{1}{2}}-(e_{S1}g_1 + e_{S2}g_2)(e_{S1}^2+e_{S2}^2)^{-\frac{3}{2}} + \mathcal{O}(g_{1,2}^2)](e_{S1}+g_1)}{\int_{-\infty}^{\infty} {\rm d} e_{S1} \int_{-\infty}^{\infty} {\rm d} e_{S2} P_S \left(e_{S1}, e_{S2}\right) [(e_{S1}^2+e_{S2}^2)^{-\frac{1}{2}}-(e_{S1}g_1 + e_{S2}g_2)(e_{S1}^2+e_{S2}^2)^{-\frac{3}{2}}+ \mathcal{O}(g_{1,2}^2)]} \nonumber\\
&\approx& g_1 \times \frac{\int_{-\infty}^{\infty} {\rm d} e_{S1} \int_{-\infty}^{\infty} {\rm d} e_{S2} P_S \left(e_{S1}, e_{S2}\right)  (e_{S1}^2+e_{S2}^2)^{-\frac{3}{2}}e_{S2}^2}{\int_{-\infty}^{\infty} {\rm d} e_{S1} \int_{-\infty}^{\infty} {\rm d} e_{S2} P_S \left(e_{S1}, e_{S2}\right) (e_{S1}^2+e_{S2}^2)^{-\frac{1}{2}}} \nonumber\\
&=& \frac{g_1}{2}
\end{eqnarray}
In this formula, we perform a Taylor expansion of the expression from the first step up to the first order in shear. Then, we eliminate the terms whose integrals are 0 by utilizing the parity of $P_S \left(e_{S1}, e_{S2}\right)$. Finally, we exploit the fact that exchanging the indices `1' and `2' would not change the result, so the numerator should be half of the denominator. Thus, we demonstrate that the SWSE method can stably estimate the shear value when $g_{1,2}$ is small.

\section{Consistency of weights between the FD test and ESD measurement}\label{app:weights}

In weak lensing, the observed ellipticity of a galaxy can be modeled as:
\begin{equation}
e_i = (1 + m_i) \gamma_i + e^s_i
\end{equation}
where $m_i$ is the multiplicative shear bias, $\gamma_i$ is the true shear, and $e^s_i$ is the intrinsic galaxy shear. Here we consider only the multiplicative bias and assume it to be identical for different shear components.
The estimator for the ESD is given by
\begin{equation}
(1 + m_A) \Delta\Sigma = \frac{\sum_i w^k \Sigma_c^k e_t^k}{\sum_i w^k},
\end{equation}
where $m_A$ is the multiplicative bias for $\Delta\Sigma$. Substituting the expression for $e^i$ and assuming $\Delta\Sigma = \Sigma_c^i \gamma_t^i$ in the absence of redshift errors, we have
\begin{equation}
(1 + m_A) \Delta\Sigma = \frac{\sum_k w^k \Sigma_c^k (1 + m^k) \gamma_t^k}{\sum_k w^k} \implies 1 + m_A = \frac{\sum_k w^k (1 + m^k)}{\sum_k w^k}
\end{equation}
in which $m^k$ denotes the multiplicative bias for the $k^{th}$ galaxy.
For the FD test, the observed ellipticity responds to the known shear $\gamma_f$ induced by field distortion. The corresponding estimator can be written as
\begin{equation}
(1 + m_B) \gamma_f
= \frac{\sum_k w^k_f e^k}{\sum_k w^k_f}.
\end{equation}
where $m_B$ is the bias measured in FD tests. It reduces to
\begin{equation}
1 + m_B = \frac{\sum_k w^k_f (1 + m^k)}{\sum_k w^k_f}.
\end{equation}
Thus, in order for the bias measured in the FD test to correctly calibrate the bias in the ESD measurement (i.e. $m_B = m_A$), the same weighting scheme must be adopted, namely $w_f^k = w^k$.

However, if redshift errors exist, the critical surface density can be written as $\tilde \Sigma_c^k = (1+\delta_c^k)\Sigma_c^k$. So the ESD bias then becomes
\begin{equation}
1 + \tilde m_A = \frac{\sum_k w^k (1 + m^k) (1 + \delta_c^k )}{\sum_k w^k} \neq \frac{\sum_k w^k (1 + m^k)}{\sum_k w^k}
\end{equation}
Therefore, photo-$z$ errors introduce additional biases that cannot be corrected by the FD test alone.

\bibliographystyle{raa}
\bibliography{bibtex}

\label{lastpage}

\end{document}